\documentclass[12pt,english]{article}

\usepackage{natbib}
\usepackage{url}
\makeatletter
\def\url@leostyle{%
    \def\UrlFont{\sf}}{\def\UrlFont{\small\ttfamily}}
\makeatother
\usepackage[font=small,labelfont=bf]{caption}

\usepackage{hyperref}

\usepackage{authblk}

\usepackage{graphicx}
\usepackage{babel}
\usepackage{amsmath}


\numberwithin{equation}{section}

\begin{document}

\title{Is Entanglement Sufficient to Enable Quantum Speedup?
}
\author{}
\date{}

\maketitle

\thispagestyle{empty}

\section{Introduction}

The mere fact that a quantum computer realises an entangled state is
ususally concluded to be insufficient to enable it to achieve quantum
computational speedup. To support this conclusion, appeal is usually
made to the Gottesman-Knill theorem
\citep[464]{nielsenChuang2000}. According to this theorem, any quantum
algorithm or protocol which exclusively utilises the elements of a
certain restricted set of quantum operations can be efficiently
simulated by classical means. Yet among the quantum computational
algorithms and informational protocols which exclusively utilise
operations from this set are some that are interesting and
important\textemdash for instance, the teleportation and superdense
coding protocols\textemdash and both of these (and others) involve the
use of \emph{entangled} quantum states. Thus \citet[]{datta2005}, for
instance, write: ``the Gottesman-Knill theorem ... demonstrates that
global entanglement is far from sufficient for exponential speedup''.

In this short note I will argue that this conclusion is misleading. As
I will explain, the quantum operations to which the Gottesman-Knill
theorem applies are precisely those which will never cause a qubit to
take on an orientation, with respect to the other subsystems
comprising the total system of which it is a part, that yields a
violation of the Bell inequalities. The fact that the Gottesman-Knill
theorem holds should therefore come as no surprise.

While it is true that more than entanglement is required to realise
quantum computational speedup in the sense that a quantum computer
implementing an entangled quantum state must utilise more than the
relatively small portion of its state space that is accessible from
the Gottesman-Knill group of transformations alone if it is to
outperform a classical computer; i.e., while it is the case that one
must \emph{use} such a state to its full potential, it is nevertheless
the case that if one is asked what \emph{physical resources} suffice
to enable one to bring about a quantum performance
advantage\textemdash a more interesting question if one is seeking for
a physical explanation for quantum speedup\textemdash then one can
legitimately answer that entanglement alone is sufficient for this
task.

\section{The Gottesman-Knill theorem}

Call an operator $A$ a \emph{stabiliser} of the state $| \psi \rangle$
if
\begin{align}
A| \psi \rangle = | \psi \rangle.
\end{align}
For instance, consider the Bell state of two qubits:
\begin{align}
| \Phi^+ \rangle = \frac{1}{\sqrt 2}(| 0 \rangle| 0 \rangle + | 1
\rangle| 1 \rangle).
\end{align}
For this state we have
\begin{align}
(X \otimes X)| \Phi^+ \rangle & = \frac{1}{\sqrt 2}(| 1 \rangle| 1
  \rangle + | 0 \rangle| 0 \rangle) \nonumber \\
  & = \frac{1}{\sqrt 2}(| 0 \rangle| 0 \rangle + | 1 \rangle| 1
  \rangle) = | \Phi^+ \rangle, \\
(Z \otimes Z)| \Phi^+ \rangle & = \frac{1}{\sqrt 2}(| 0 \rangle| 0
  \rangle + (-| 1 \rangle)(-| 1 \rangle)) \nonumber \\
  & = \frac{1}{\sqrt 2}(| 0 \rangle| 0 \rangle + | 1 \rangle| 1
  \rangle) = | \Phi^+ \rangle.
\end{align}
$X \otimes X$ and $Z \otimes Z$ are thus both stabilisers of the state
$| \Phi^+ \rangle$. Here, $X$ and $Z$ are the Pauli operators:
\begin{align}
  X \equiv \sigma_1 \equiv \sigma_x \equiv
  \left(
  \begin{matrix}
  0 & 1 \\
  1 & 0
  \end{matrix}
  \right), &
  \quad Z \equiv \sigma_3 \equiv \sigma_z \equiv
  \left(
  \begin{matrix}
  1 & 0 \\
  0 & -1
  \end{matrix}
  \right).
\end{align}
The remaining Pauli operators, $I$ (the identity operator) and $Y$,
are defined as:
\begin{align}
  I \equiv \sigma_0 \equiv \sigma_I \equiv
  \left(
  \begin{matrix}
  1 & 0 \\
  0 & 1
  \end{matrix}
  \right), &
  \quad Y \equiv \sigma_2 \equiv \sigma_y \equiv
  \left(
  \begin{matrix}
  0 & -i \\
  i & 0
  \end{matrix}
  \right).
\end{align}
The Pauli group, $P_n$, of $n$-fold tensor products of Pauli operators
(for instance, for $n = 2$, $P_2 \equiv \{I \otimes I, I \otimes X, I
\otimes Y, I \otimes Z, X \otimes I, X \otimes X, X \otimes Y,
... \}$) is an example of a group of operators closed under matrix
multiplication.

Call the set, $V_S$, of states that are stabilised by every element in
$S$, where $S$ is some group of operators closed under matrix
multiplication, the \emph{vector space stabilised by} $S$. Consider a
state $| \psi \rangle \in V_S$. From the definition of a unitary
operator, we have, for any $s \in S$ and any unitary operation $U$,
\begin{align}
U| \psi \rangle = Us| \psi \rangle = UsU^\dagger U| \psi \rangle.
\end{align}
Thus $UsU^\dagger$ stabilises $U| \psi \rangle$ and the vector space
$UV_S$ is stabilised by the group $USU^\dagger \equiv \{UsU^\dagger|s
\in S\}$. Consider, for instance, the state $| 0 \rangle$, stabilised
by the $Z$ operator. To determine the stabiliser of this state after
it has been subjected to the (unitary) Hadamard transformation $H| 0
\rangle = | + \rangle$ we simply compute $HZH^\dagger$. Thus the
stabiliser of $| + \rangle$ is $X$.

Now let $s_1, ..., s_n$ be elements of $S$. $s_1, ..., s_n$ are said
to \emph{generate} the group $S$ if every element of $S$ can be
written as a product of elements from $s_1, ..., s_n$. For instance,
the reader can verify that the subgroup, $A$, of $P_3$, defined by $A
\equiv \{I^{\otimes 3}, Z \otimes Z \otimes I, I \otimes Z \otimes Z, Z
\otimes I \otimes Z\}$ can be generated by the elements $\{Z \otimes Z
\otimes I, I \otimes Z \otimes Z\}$ \citep[\textsection
  10.5.1]{nielsenChuang2000}. We may thus alternately express $A$ in
terms of its generators as follows: $A = \langle Z \otimes Z \otimes
I, I \otimes Z \otimes Z \rangle$.

In order to compute the action of a unitary operator on a group $S$ it
suffices to compute the action of the unitary operator on the
generators of $S$. For instance, $| 0 \rangle^{\otimes n}$ is the
unique state stabilised by $\langle Z_1, Z_2, ..., Z_n\rangle$ (where
the latter expression is a shorthand form of $\langle Z \otimes
I^{\otimes   n-1}, I \otimes Z \otimes I^{\otimes n-2}, ...,I^{\otimes
  n-1} \otimes Z\rangle$). Consequently, the stabiliser of the state
$H^{\otimes n}| 0 \rangle^{\otimes n}$ is $\langle X_1, X_2, ...,
X_n\rangle$. Note that this state, expressed in the standard state
vector formalism,
\begin{align}
H^{\otimes n}| 0 \rangle^{\otimes n} & = \left(\frac{1}{2^{n/2}}(| 0
\rangle + | 1 \rangle)^n \right ) \nonumber \\
& = \left (\frac{1}{2^{n/2}}\sum_x^{2^n-1}| x \rangle \right),
\end{align}
specifies $2^n$ different amplitudes. Contrast this with the
stabiliser description of the state in terms of its generators
$\langle X_1, X_2, ..., X_n\rangle$, which is linear in $n$ and thus
capable of an efficient classical representation.

It turns out that, using the stabiliser formalism, all (as well as all
combinations) of the following gates are capable of an efficient
classical representation: \emph{Pauli gates, Hadamard gates, phase
  gates (i.e.,$\pi/2$ rotations of the Bloch sphere for a single qubit
  about the $\hat{z}$-axis), and CNOT gates; as well as state
  preparation in the computational basis and measurements of the Pauli
  observables.} This is the content of the \emph{Gotteman-Knill
  theorem} \citep[\textsection 10.5.4]{nielsenChuang2000}.

In fact, many important quantum algorithms utilise gates from this set
of operations exclusively. One of these, for instance, is the
well-known teleportation algorithm \citep[cf.,][]{bennett1993}. But
what is especially notable about this theorem from the point of view
of our discussion is that some of the states which may be realised
through the operations in this set are actually entangled states. In
particular, by combining a Hadamard and a CNOT gate, one can generate
any one of the Bell states (which one is generated depends on the
value assigned to the input qubits); i.e.,
\begin{align}
| 0 \rangle| 0 \rangle \xrightarrow{H \otimes I} \frac{| 0 \rangle| 0
\rangle + | 1 \rangle| 0 \rangle}{\sqrt 2} \xrightarrow{CNOT} \frac{|
  0 \rangle| 0 \rangle + | 1 \rangle| 1 \rangle}{\sqrt 2} = | \Phi^+
\rangle, \\
| 0 \rangle| 1 \rangle \xrightarrow{H \otimes I} \frac{| 0 \rangle| 1
\rangle + | 1 \rangle| 1 \rangle}{\sqrt 2} \xrightarrow{CNOT} \frac{|
  0 \rangle| 1 \rangle + | 1 \rangle| 0 \rangle}{\sqrt 2} = | \Psi^+
\rangle, \\
| 1 \rangle| 0 \rangle \xrightarrow{H \otimes I} \frac{| 0 \rangle| 0
\rangle - | 1 \rangle| 0 \rangle}{\sqrt 2} \xrightarrow{CNOT} \frac{|
  0 \rangle| 0 \rangle - | 1 \rangle| 1 \rangle}{\sqrt 2} = | \Phi^-
\rangle, \\
| 1 \rangle| 1 \rangle \xrightarrow{H \otimes I} \frac{| 0 \rangle| 1
\rangle - | 1 \rangle| 1 \rangle}{\sqrt 2} \xrightarrow{CNOT} \frac{|
  0 \rangle| 1 \rangle - | 1 \rangle| 0 \rangle}{\sqrt 2} = | \Psi^-
\rangle.
\end{align}
In fact many quantum algorithms utilise just such a combination of
gates. If all of the operations from this set are efficiently
classically simulable, however, then it appears as though
entanglement, by itself, cannot be a sufficient resource for realising
quantum speedup, for evidently there are quantum algorithms utilising
entangled states that are efficiently simulable classically.

In what follows I will argue that this conclusion is not warranted. An
entangled state does, in fact, provide sufficient resources to enable
quantum computational speedup. What the Gottesman-Knill theorem
actually shows is not that entanglement is insufficient, but that (not
surprisingly) it is possible to utilise the resource provided by an
entangled state to less than its full potential.

\section{Bell's theorem}

For a system in the singlet state ($| \Psi^- \rangle$), joint
experiments on its subsystems are related by the following expression
for the expectation value of these combined experiments:
\begin{align}
\label{rol:eqn:singlet}
\langle \sigma_m \otimes \sigma_n \rangle = - \hat{m} \cdot \hat{n} =
- \cos\theta.
\end{align}
Here $\sigma_m, \sigma_n$ represent spin-$m$ and spin-$n$ experiments
on the first (Alice's) and second (Bob's) system, respectively, with
$\hat{m}, \hat{n}$ the unit vectors representing the orientations of
the two experimental devices, and $\theta$ the difference in these
orientations. Note, in particular, that when $\theta = 0$, $\langle
\sigma_m \otimes \sigma_n \rangle = -1$ (i.e., experimental results
for the two subsystems are perfectly anti-correlated), when $\theta =
\pi$, $\langle \sigma_m \otimes \sigma_n \rangle = 1$ (i.e.,
experimental results for the two subsystems are perfectly correlated),
and when $\theta = \pi/2$, $\langle \sigma_m \otimes \sigma_n \rangle
= 0$ (i.e., experimental results for the two subsystems are not
correlated at all).

Consider the following attempt \citep[]{bell1964} to reproduce the
quantum mechanical predictions for this state by means of a hidden
variables theory. Let the hidden variables of the theory assign, at
state preparation, to each subsystem of a bipartite quantum system, a
unit vector $\hat{\lambda}$ (the same value for $\hat{\lambda}$ is
assigned to each subsystem) which determines the outcomes of
subsequent experiments on the system as follows. Let the functions
$A_\lambda(\hat{m}), B_\lambda(\hat{n})$ represent, respectively, the
outcome of a spin-$m$ and a spin-$n$ experiment on Alice's and Bob's
subsystem. Define these as:
\begin{align}
A_\lambda(\hat{m}) & = \mbox{sign} (\hat{m} \cdot \hat{\lambda}),
\nonumber \\
B_\lambda(\hat{n}) & = - \mbox{sign} (\hat{n} \cdot \hat{\lambda}),
\end{align}
where $\mbox{sign}(x)$ is a function which returns the sign (+, -) of
its argument.

The reader can verify that the probability that both
$A_\lambda(\hat{m})$ and $B_\lambda(\hat{n})$ yield the same value,
and the probability that they yield values that are different
(assuming a uniform probability distribution over $\hat{\lambda}$),
are respectively:
\begin{align}
\mbox{Pr}(+, +) & = \mbox{Pr}(-, -) = \theta/2\pi, \nonumber \\
\mbox{Pr}(+, -) & = \mbox{Pr}(-, +) = \frac{1}{2}\left(1 -
\frac{\theta}{\pi}\right),
\end{align}
with $\theta$ the (positive) angle between $\hat{m}$ and $\hat{n}$. This
yields, for the expectation value of experiments on the combined state:
\begin{align}
\langle \sigma_m \otimes \sigma_n \rangle = \frac{2 \theta}{\pi} - 1.
\end{align}

When $\theta$ is a multiple of $\pi/2$, this expression yields
predictions identical to the quantum mechanical ones: perfect
anti-correlation for $\theta \in \{0, 2\pi, ...\}$, no correlation for
$\theta \in \{\pi/2, 3\pi/2, ...\}$, and perfect correlation for $\theta
\in \{\pi, 3\pi, ...\}$. However, for all other values of $\theta$ there
are divergences from the quantum mechanical predictions.

It turns out that this is not a special characteristic of the simple
hidden variables theory considered above. \emph{No} hidden variables
theory is able to reproduce the predictions of quantum mechanics if it
makes the very reasonable assumption that the probabilities of local
experiments on Alice's subsystem (and likewise Bob's) are completely
determined by Alice's local experimental setup together with a hidden
variable taken on by the subsystem at the time the joint state is
prepared. Consider the following expression relating different spin
experiments on Alice's and Bob's respective subsystems for arbitrary
directions $\hat{m}, \hat{m}', \hat{n}, \hat{n}'$:
\begin{align}
| \langle \sigma_m \otimes \sigma_n \rangle + \langle \sigma_m \otimes
\sigma_{n'} \rangle | + | \langle \sigma_{m'} \otimes \sigma_n \rangle
- \langle \sigma_{m'} \otimes \sigma_{n'} \rangle |.
\end{align}
As before, let $A_\lambda(\hat{m}) \in \{\pm 1\}, B_\lambda(\hat{n})
\in \{\pm 1\}$ represent the results, given a specification of some
hidden variable $\lambda$, of spin experiments on Alice's and Bob's
subsystems. We make no assumptions about the nature of the `common
cause' $\lambda$ this time\textemdash it may take any form. What we do
assume is that, as I mentioned above, the outcomes of Alice's
experiments depend only on her local setup and on the value of
$\lambda$; i.e., we do not assume any further dependencies between
Alice's and Bob's local experimental configurations. This allows us to
substitute $\langle A_\lambda(\hat{m}) \cdot B_\lambda(\hat{n})
\rangle$ for $\langle \sigma_m \otimes \sigma_n \rangle$, thus
yielding:
\begin{align}
\label{rol:eqn:chsh1}
& \big | \big\langle A_\lambda(\hat{m})B_\lambda(\hat{n}) \big\rangle +
  \big\langle A_\lambda(\hat{m})B_\lambda(\hat{n}') \big\rangle \big | +
  \big | \big\langle A_\lambda(\hat{m}')B_\lambda(\hat{n}) \big\rangle -
  \big\langle A_\lambda(\hat{m}')B_\lambda(\hat{n}') \big\rangle \big |
  \nonumber \\
& = \big | \big\langle A_\lambda(\hat{m})\big(B_\lambda(\hat{n}) +
  B_\lambda(\hat{n}')\big)\big\rangle \big | + \big | \big\langle
  A_\lambda(\hat{m}')\big(B_\lambda(\hat{n}) -
  B_\lambda(\hat{n}')\big)\big\rangle \big | \nonumber \\
& \leq \big\langle \big | A_\lambda(\hat{m})\big(B_\lambda(\hat{n}) +
  B_\lambda(\hat{n}')\big)\big | \big\rangle + \big\langle \big |
  A_\lambda(\hat{m}')\big(B_\lambda(\hat{n}) - B_\lambda(\hat{n}')\big)
  \big | \big\rangle,
\end{align}
which, since $|A_\lambda(\cdot)| = 1$, is
\begin{align}
& \leq \big\langle \big | B_\lambda(\hat{n}) + B_\lambda(\hat{n}')\big |
\big\rangle + \big\langle  \big | B_\lambda(\hat{n}) -
B_\lambda(\hat{n}') \big | \big\rangle \nonumber \\
& \leq 2,
\end{align}
where the last inequality follows from the fact that $B_\lambda(\cdot)$
can only take on values of $\pm 1$. This expression, a variant of the
`Bell inequality' \citeyearpar[]{bell1964}, is known as the
\emph{Clauser-Horne-Shimony-Holt} (CHSH) inequality
\citep[cf.,][]{chsh1969,bell1981}.

Quantum mechanics violates the CHSH inequality for some experimental
configurations. For example, let the system be in the singlet state;
i.e., such that its statistics satisfy \eqref{rol:eqn:singlet}; and let
the unit vectors $\hat{m}, \hat{m}', \hat{n}, \hat{n}'$ (taken to lie
in the same plane) have the orientations $0, \pi/2, \pi/4, -\pi/4$
respectively. The differences, $\theta$, between the different
orientations (i.e., $\hat{m} - \hat{n}, \hat{m} - \hat{n}', \hat{m}' -
\hat{n}$, and $\hat{m}' - \hat{n}'$) will all be in multiples of
$\pi/4$ and we will have:
\begin{align}
\langle \sigma_m \otimes \sigma_n \rangle & =
\langle \sigma_m \otimes \sigma_{n'} \rangle =
\langle \sigma_{m'} \otimes \sigma_n \rangle = \sqrt 2/2, \\
\langle \sigma_{m'} \otimes \sigma_{n'} \rangle & = -\sqrt 2/2, \\
| \langle \sigma_m \otimes \sigma_n \rangle & + \langle \sigma_m \otimes
\sigma_{n'} \rangle | + | \langle \sigma_{m'} \otimes \sigma_n \rangle
- \langle \sigma_{m'} \otimes \sigma_{n'} \rangle | = 2\sqrt 2
\not\leq 2.
\end{align}

The predictions of quantum mechanics for arbitrary orientations
$\hat{m}, \hat{m}', \hat{n}, \hat{n}'$ cannot, therefore, be
reproduced by a hidden variables theory in which all correlations
between subsystems are due to a common parameter endowed to them at
state preparation. They can, however, be reproduced by such a hidden
variables theory for certain special cases. In particular, the
inequality is \emph{satisfied} (as the reader can verify) when
$\hat{m}$ and $\hat{n}$, $\hat{m}$ and $\hat{n}'$, $\hat{m}'$ and
$\hat{n}$, and $\hat{m}'$ and $\hat{n}'$ are all oriented at angles
with respect to one another that are given in multiples of $\pi/2$.

\section{Entanglement as a sufficient resource}

Recall the content of the Gottesman-Knill theorem: \emph{Pauli gates,
  Hadamard gates, phase gates, and CNOT gates; as well as state
  preparation in the computational basis and measurements of the Pauli
  observables} are efficiently simulable by a classical computer. It
is commonly concluded, from this, that entanglement cannot therefore
be sufficient to enable a quantum algorithm to achieve a speedup over
its classical counterpart. When one notes that all of the operations
which comprise this set involve rotations of the Bloch sphere that are
multiples of $\pi/2$, however, the fact that algorithms restricted to
just these operations are classically simulable should come as no
surprise. In a multi-partite system, no amount of $k\pi/2$
transformations of one of the constituent qubits will cause it to take
on an orientation with respect to the other subsystems that is not a
multiple of $\pi/2$ (unless it was so oriented initially). And as we
have seen above, the statistics of compound states for which the
difference in orientation between subsystems is a multiple of $\pi/2$
are capable in general of being reproduced by a classical hidden
variables theory.

In light of this it is misleading, I believe, to conclude, on the
basis of the Gottesman-Knill theorem, that entanglement is not a
sufficient resource to enable quantum computational speedup. What the
Gottesman-Knill theorem shows us is that simply having an entangled
state is not enough to enable one to outperform a classical computer;
one must also \emph{use} such a state to its full advantage; i.e., one
must not limit oneself to transformations which utilise only a small
portion of the system's allowable state space. In this sense, it is
indeed correct to say that entanglement is insufficient to enable
quantum speedup. However, if one intends by the claim that
entanglement is insufficient\textemdash something very
different\textemdash that \emph{further physical resources} are
required to enable speedup, then I submit that this claim is
incorrect.

It is possible to characterise the distinction between classical and
quantum mechanical systems as follows. Whereas the nature of quantum
mechanical systems is such that they allow us to fully exploit the
representational capacity of Hilbert space, this is not so for
classical systems \citep[cf.,][]{ekert1998}. Thus the state space of an
$n$-fold quantum system that is efficiently simulable classically is
only a tiny portion of the system's overall state space. The reason
for the larger size of a quantum state space, however, is the
possibility of entangled quantum systems. It is because composite
classical systems must always be representable by product states that
their state space is smaller. If we have an $n$-fold entangled quantum
system, therefore, it follows straightforwardly that such a system
cannot, \emph{in general}, be simulated classically.

Evidently, it is possible to utilise only a small portion of the state
space of such a system\textemdash exactly that portion of the state
space that is accessible efficiently by an $n$-fold classical
system\textemdash but this has no bearing on the nature of the actual
physical resources that are provided by the quantum
system. Analogously, a life vest may be said to be sufficient to keep
me afloat on liquid water. I must actually wear it if it is to perform
this function, of course; but that is not a fact about this piece of
equipment's capabilities, only about my choice whether to use it or
not.

What if the waves are rough? It may be that in this case my life vest
will not be sufficient to save me. Analogously, in the presence of
noise, as noted by \citet[]{linden2001}, entanglement may not be
sufficient to enable one to achieve \emph{exponential} quantum
speedup. Nevertheless, even in rough weather I will at least have a
better chance of surviving with my life vest on than I will without
it. Likewise, even in the presence of noise, an entangled quantum
state will be sufficient to enable some (though perhaps only a
sub-exponential) quantum speedup.

Far from being a problem for the view that entanglement is a
sufficient resource to enable quantum speedup, the Gottesman-Knill
theorem serves to \emph{highlight} the role that is actually played by
entanglement in the quantum computer and to clarify exactly why and in
what sense it is sufficient.

\bibliographystyle{apa-good}
\bibliography{Bibliography}{}

\end{document}